\documentclass[preprint,3p,12pt]{elsarticle}
\usepackage{amssymb}
\usepackage{lineno,hyperref}
\usepackage{graphicx}
    \makeatletter
    \def\ps@pprintTitle{%
      \let\@oddhead\@empty
      \let\@evenhead\@empty
      \def\@oddfoot{\reset@font\hfil\thepage\hfil}
      \let\@evenfoot\@oddfoot
    }
    \makeatother

\begin{document}

\begin{frontmatter}

\title{Does the hierarchy problem generate the seesaw scale?} %\tnoteref{mytitlenote}}

\author{Ravi Kuchimanchi}%\fnref{myfootnote}}
%\address{19 Idlewild Street, Bel Air, MD 21014.}


\cortext[mycorrespondingauthor]{Corresponding author}
\ead{raviaravinda@gmail.com}/


%\begin{abstract}
%This template helps you to create a properly formatted \LaTeX\ manuscript.
%\end{abstract}

\begin{abstract}

We find that minimizing the number of fine tuning relations in non-supersymmetric models can determine the scales at which some gauge symmetries beyond the standard model must break. We show that $SU(2)_R$ and  $B-L$ gauge symmetries of the minimal left-right symmetric model must break at a scale $10^{15} GeV$ or higher, determined by the hierarchy problem and small ratios of quark masses, if parameters that break chiral or $\mu-$symmetries (and therefore can be naturally small), are not fine-tuned. This provides the $raison \ d'\hat{e}tre$ for the seesaw scale $\sim 10^{15}GeV$ indicated by neutrino experiments. Small ratios of fermion (quark) masses, which  are natural in the standard model due to approximate chiral symmetry, will have to be fine tuned in minimal left right model if $SU(2)_R \times U(1)_{B-L}$ breaks at a lower scale.

\end{abstract}
\begin{keyword}
\texttt{Hierarchy problem\sep fine tuning \sep left-right symmetry}
\PACS  12.10.Kt \sep 11.15.Ex \sep 11.30.Rd \sep 14.60.St
%\MSC[2010] 00-01\sep  99-00
\end{keyword}

\end{frontmatter}

%\linenumbers

\section{Introduction:} 
What we make of the hierarchy  between the  weak scale ($v_{wk} \sim 246 GeV$) and the reduced Planck scale ($M_{Pl} \sim 2.4 \times 10^{18} GeV$), can determine where we expect the next scale of new physics to be. 
%Since setting $v_{wk}$ to zero does not restore any symmetries in the standard model, this hierarchy cannot be  due to a small breaking of a chiral symmetry.  
Note that radiative corrections to $v_{wk}^2$ of the order $h_t^2 M_{Pl}^2 / (16 \pi^2)$ (with top Yukawa coupling $h_t \sim 1$) would shift it towards the Planck scale unless there is fine tuning or a precise cancellation of such large terms to generate the small weak scale.  This fine tuning of the standard model is the well known  hierarchy problem~\cite{PhysRevD.14.1667,PhysRevD.19.1277,PhysRevD.20.2619}. 

In order to avoid fine tuning, new physics such as supersymmetry that solves the hierarchy problem was expected close to the weak scale. %However several precision experiments such as efforts to detect electron and neutron EDMs, flavour changing currents as well as direct searches at the LEP, Tevatron and LHC (running at 8 TeV) have found no such new physics.
However despite direct and indirect searches, no hint for such new physics has been found. Thus the minimal supersymmetric standard model  is now considered to be fine tuned to a 1\% level or worse (see for example~\cite{Arvanitaki:2013yja,Bertuzzo:2013dja,Anton}).  It appears that nature accepts fine tuning of gauge symmetry breaking scales and it is possible that there is no supersymmetry (or no SUSY till Planck scale). 

%However there may be other intermediate symmetry breaking scales between the weak and Planck scale.  
However there is some evidence for a new high energy scale in nature.  The neutrino mass data~\cite{Beringer:1900zz} ($|\Delta m^2_{32}| = 0.0023 eV^2$) points to a seesaw~\cite{Minkowski:1977sc,van1979supergravity,Yanagida01091980,Mohapatra:1979ia} scale  $M_{ss} = v_{wk}^2 / \sqrt{|\Delta m^2_{32}|} \sim 1.3 \times 10^{15} GeV$. But what is the $raison \ d'\hat{e}tre$ for  this scale? %Why is it so much higher than the weak scale? %which is well below the Planck scale?  % The see-saw mechanism violates Lepton number ($B-L$ symmetry) by providing large Majorana masses to the right handed neutrinos. Thus gauge forces such as $U(1)_{B-L}$, $SU(2)_R$ and even grand unification that may exist in nature, must be broken at or above the seesaw scale. The scales of breaking of these gauge symmetries would need fine-tuning.

%The standard lore is that such a scale is due to grand unification -- if three couplings, the strong, weak and electromagnetic unify through renormalization group running, there are then two relations between the couplings that determine the unification scale (such as $M_X$ of $SO(10)$ breaking) and an intermediate scale (such as $v_R$ or $v_{B-L}$ where a sub-group of $SO(10)$ breaks). Since the see-saw mechanism breaks Lepton number (or $B-L$ symmetry) the seesaw scale is identified with the scale $v_{B-L}$. 

The standard lore is that this is near where the weak, strong and electro-magnetic forces unify.   However several grand unified theories (GUTs) are tightly constrained or ruled out. For example a recent paper~\cite{Altarelli:2013aqa} finds that only one chain of non-supersymmetric $SO(10)$ breaking is still consistent with data and that the constraint of unification of couplings determines the intermediate  seesaw scale to be $\sim 10^{11} GeV$, with grand unification at $10^{16} GeV$.  However the model suffers from a ``very large level of fine-tuning" (quoting~\cite{Altarelli:2013aqa}) and ``...the idea of an $SO(10)$ GUT is very appealing but all its practical realizations are clumsy, more  so in the non SUSY case because of the hierarchy problem...." . 

%In theories with several Higgs multiplets and gauge scales, quadratic divergences imply that several parameters must be fine-tuned. 
 Historically the \emph{extended survival hypothesis}~\cite{delAguila:1980at,Dimopoulos:1984ha,PhysRevD.27.1601} used in grand unified models such as~\cite{Altarelli:2013aqa} and other extensions of the standard model such as the left-right symmetric model~\cite{PhysRevD.27.1601}, was motivated by the desire to minimize the number of fine tuning relations~\cite{Dimopoulos:1984ha,PhysRevD.27.1601}. The \emph{hypothesis} however is to minimize the Higgs multiplets that are needed at lower and intermediate mass scales, rather than the number of fine tunings.  It is implicitly assumed that by having the minimal Higgs content at lower scales, the number of fine tunings are automatically minimized \footnote{Quoting from Dimopoulos and Georgi \cite{Dimopoulos:1984ha}``\emph{3.The extended survival hypothesis.}  If supersymmetry is not relevant (either absent completely or broken at the largest scale), we can still keep scalars light, but we need one fine tuning of parameters for each light scalar multiplet.  Under these circumstances, we may want to keep the number of light scalars and therefore the number of independent fine tunings  to the absolute minimum necessary."  }. However in practice since we do not keep track of the number of fine tunings, it may not always turn out to be minimal, and the theory becomes less predictive.    %Therefore  minimization of the  number of fine tunings has to be more carefully looked at, and cannot be assumed to follow only from the choice of the minimal Higgs content. 

%However we note that if an arbitrary number of  fine tunings that keep the Higgs multiplet masses at their hypothesized scales are permitted, they may not always automatically be minimal. Therefore  minimization of the  number of fine tunings has to be more carefully looked at, and cannot be assumed to follow only from the choice of the minimal Higgs content. 
%Since its not just the minimal number of vacuum expectation values (VEVs) needed for breaking gauge symmetries that are fine tuned, information is lost about what the hierarchy problem itself may be contributing to a particular scale due to small fermion mass ratios that  break chiral symmetries, which are usually required to obtain a particular symmetry breaking pattern. 

Therefore we  ask, how many fine tuning constraints are actually necessary in non-supersymmetric theories to keep the scale of  gauge symmetry breaking small?  In the standard model $SU(2)_L \times U(1)_Y$ breaks to $U(1)_{em}$ requiring  one Higgs vacuum expectation value (VEV) that is kept small by fine-tuning.  If the standard model is extended to the left-right symmetric model based on $SU(2)_L \times SU(2)_R \times U(1)_{B-L}$, then we can first break $SU(2)_R \times U(1)_{B-L}$ to $U(1)_Y$ at the scale $v_R \sim v_{B-L}$.  In exact analogy with the standard model we expect that one fine-tuning relation is required to keep $v^2_R \ll M^2_{Pl}$.  Therefore, totally we require two fine tunings for gauge scales in this model - one to ensure $v^2_R \ll M^2_{Pl}$ and the other to ensure $v^2_{wk} \ll M^2_{Pl}$.    

In this letter we show that if we allow exactly two fine-tuning relations in the minimal left-right symmetric model,  while  $v^2_R  \ll M^2_{Pl}$ is possible, $v_R$ cannot be kept arbitrarily small and there is a meaningful lower bound $v_R \sim v_{B-L} \gtrsim 10^{15} GeV$. The reason for the bound is that a chiral $\mu-$symmetry is needed along with the two fine-tuning relations to obtain the correct symmetry breaking pattern.  However the $\mu-$symmetry is approximately broken by small second  generation Yukawa terms to obtain the proper quark mass spectrum.  
%The hierarchy problem then re-surfaces but at a scale $\sim 10^{15}GeV$ that is suppressed from the Planck scale by the second generation Yukawas, which leads to the above lower bound on $v_R \sim v_{B-L}$.   
The hierarchy problem then reappears to destabilize the symmetry breaking pattern, but  the  quadratic divergence from the Planck scale is now suppressed by second generation Yukawas, leading to a scale of $10^{15} GeV$ for $v_R \sim v_{B-L}$. % If $v_R \sim v_{B-L}$ is below this scale, the left-right symmetric model does not break to the standard model, but breaks to a different group. 
Breaking of $SU(2)_R \times U(1)_{B-L}$ triggers the seesaw mechanism, and the  see-saw scale $\sim 10^{15} GeV$ hinted at by neutrino mass data is thus generated by the hierarchy problem. % If our bound is to be avoided a third fine-tuning (that of the chiral $\mu-$symmetry breaking parameter) is required in the left-right symmetric model. 

Thus without using any grand unification constraints, we can obtain meaningful bounds on gauge symmetry breaking scales and a $raison \ d'\hat{e}tre$ for the seesaw scale. The idea of minimizing the number of fine-tunings and using the hierarchy problem and chiral symmetries to provide limits on gauge symmetry breaking scales can be generalized to other groups.

\section{Hierarchy problem in left-right model}
We consider the minimal Left-Right symmetric model~\cite{PhysRevD.10.275,PhysRevD.11.566,Senjanovic:1975rk} based on $G_{LR} \equiv SU(3)_c \times SU(2)_L \times SU(2)_R \times U(1)_{B-L}$, with the minimal number of scalar fields needed to break the gauge symmetries,  namely triplets $\Delta_R$ (1, 1, 3, 2) and $\Delta_L$ (1, 3, 1, 2), and bi-doublet $\phi$ (1, 2, 2, 0). Note that $\Delta_L$ is needed  since there is a space-inversion or parity (P) symmetry under which, as is usual in left-right  symmetric (LR) models, the space-time coordinates $(x,t) \rightarrow (-x,t), \ \phi \rightarrow \phi^\dagger$ and subscripts $L \leftrightarrow R$ for all other fields (see for example~\cite{Duka:1999uc}).   The scalar fields have the form
\begin{equation}
\begin{array}{cc}
\phi = \left(\begin{array}{cc}
\phi^o_1 & \phi^+_2 \\
\phi^-_1 & \phi^o_2
\end{array}
\right),  &
\Delta_{L,R} = \left(\begin{array}{cc}
\delta^+_{L,R} / \sqrt{2} & \delta^{++}_{L,R} \\
\delta^o_{L,R} & - \delta^+_{L,R} /\sqrt{2}
\end{array}
\right).
\end{array}
%\nonumber
\label{eq:fields}
\end{equation}
The VEV $\left<\delta^o_R\right>$ breaks $SU(2)_R \times U(1)_{B-L}$ to $U(1)_Y$, while $\left<\phi_{1,2}^o\right>$ cause electro-weak symmetry breaking, and $\left<\delta^o_L\right>$ is a much smaller induced VEV. We designate these by
\begin{equation}
\label{eq:vev}
\left<\delta^o_R\right> = {v_R \over  \sqrt{2}}, \  \left<\phi_1^o\right>= {k_1 \over \sqrt{2}}, \ \left<\phi_2^o\right>= {k_2 \over \sqrt{2}}, \ \left<\delta^o_L\right> = {v_L \over \sqrt{2}}
\end{equation}  
with $v^2_{wk} = |k_1|^2 + |k_2|^2$ and study the fine-tuning implications to obtain the hierarchy $v^2_{wk} << v_R^2 << M^2_{Pl}$, where the cut-off scale of the theory is taken to be $M_{Pl}$.  Note that the   fine tuning of the weak scale from $v_R$ scale in left-right models was discussed in~\cite{PhysRevD.65.095003}.  Likewise fine tuning of the weak scale in the standard model from the seesaw scale was discussed in~\cite{Vissani:1997ys,Casas:2004gh}.  However fine tuning issues of weak and $v_R$ scales due to quadratically divergent radiative corrections from a cut-off scale (such as $M_{Pl}$) much greater than $v_R$ or seesaw scale were not previously studied.

To simplify calculations, without loss of generality, we take all  VEVs to be real.  In fact all parameters of the Higgs potential are real due to parity, except for one that is discussed in the comments at the end.% as the only $P-$invariant complex quartic coupling breaks $\mu-$symmetry.

Relevant Higgs potential terms (using standard notation, see for example~\cite{Duka:1999uc,PhysRevD.65.095003})  that involve only $\Delta_R$ are
\begin{eqnarray}
-\mu_3^2 Tr (\Delta_R^\dagger \Delta_R) + \rho_1 [Tr(\Delta_R^\dagger \Delta_R)]^2 +  \rho_2 [Tr(\Delta_R \Delta_R) Tr(\Delta_R^\dagger \Delta_R^\dagger)]
\end{eqnarray}

Substituting for the VEV $v_R$ from equations~(\ref{eq:vev}) and~(\ref{eq:fields}), and ignoring coupling terms with $\phi$, the above potential can be minimized (for $\mu^2_3, \rho_1, \rho_2 > 0$) to give $v_R^2 = \mu^2_3 / \rho_1$
%\begin{equation}\label{eq:v_R} 
%v_R^2 = {\mu^2_3 \over \rho_1} 
%\end{equation}
Since $\mu^2_3$ receives radiative correction of the order $M_{Pl}^2$ owing to the hierarchy problem, it is understood that it has been fine tuned to cancel these corrections, so that $v^2_R$ can be much less than $M^2_{Pl}$.  This fine tuning cannot be avoided in non-supersymmetric models 

We  impose a chiral $\mu-$symmetry under which $\phi \rightarrow e^{i\beta} \phi$ and $\Delta_R$ is invariant (we provide transformations of other fields when they appear later),  and write  the terms involving $\phi$ (but ignore terms involving $\Delta_L$ until later) in the Higgs potential responsible for electro-weak symmetry breaking: 
\begin{eqnarray}
-\mu_1^2 Tr(\phi^\dagger \phi) + \lambda_1 [Tr(\phi^\dagger \phi)]^2  + \lambda_3 Tr(\tilde{\phi}^\dagger \phi) Tr(\tilde{\phi}\phi^\dagger) + \nonumber \\  \alpha_1 Tr(\phi^\dagger \phi) Tr(\Delta_R^\dagger \Delta_R) + \alpha_3 Tr (\phi^\dagger \phi \Delta_R \Delta_R^\dagger)
\end{eqnarray} 
where $\tilde{\phi} = \tau_2 \phi^\star \tau_2$. As we shall see later, the $\mu-$symmetry is needed to obtain the correct symmetry breaking pattern to the standard model.
Substituting for the VEVs from equation~(\ref{eq:vev}) we can rewrite the above as
\begin{eqnarray}
\left(-\mu_1^2 +{\alpha_1 \over 2} v_R^2 \right) {k_1^2 \over 2} + \left(-\mu_1^2 + {\alpha_1 \over 2} v_R^2 +  {\alpha_3 \over 2} v_R^2 \right) {k_2^2 \over 2} + \nonumber \\ {\lambda_1 \over 4}\left(k_1^4 + k_2^4\right) + \left({\lambda_1 \over 2} + \lambda_3\right) k_1^2k_2^2 
\label{eq:higgskk'}
\end{eqnarray}
It is easy to see that if $\mu_1^2$ is fine tuned so that the quantity in the brackets of the first term of equation~(\ref{eq:higgskk'}) is of the order of the $-v_{wk}^2$, minimization with respect to $k_2$ and $k_1$ leads to $k_2=0$ and $v_{wk} \approx \ k_1  = \sqrt{(\mu_1^2 - \alpha_1 v_R^2 / 2) / \lambda_1}$.  This is the usual fine tuning that needs to be done to keep the weak scale small. %It is implicit that $\mu_1^2$ has been tuned to not only cancel  $\alpha_1 v_R^2/2$  up to the weak scale, but also to cancel the quadratically divergent radiative corrections to itself. 

Note that we have assumed $\alpha_3 > 0$ and the quantity in brackets of second term of eq.~(\ref{eq:higgskk'}) can be rewritten as $(\alpha_3 /2) v_R^2 + O(v_{wk}^2) \sim (\alpha_3 /2) v_R^2$.  This implies that the second Higgs doublet (that is fields with subscript 2 in matrix representing bidoublet $\phi$ in equation~(\ref{eq:fields})) has a mass $m_{H_2} = \left(\sqrt{{\alpha_3 / 2}}\right)v_R$.
%\begin{equation} 
%m_{H_2} = \left(\sqrt{{\alpha_3 \over 2}}\right)v_R \label{eq:second}
%\end{equation}.

We can in principle provide a small VEV to $k_2$ without any more fine tuning by breaking the $\mu-$symmetry softly using the term,
\begin{equation} V_{break} = - \mu_2^2 Tr (\tilde{\phi}^\dagger\phi) +h.c.
\label{eq:vbreak}
\end{equation} 
This term adds $-2 \mu_2^2 k_1 k_2 $ to  equation~(\ref{eq:higgskk'}). Minimizing with respect to $k_2$ now gives to the lowest order \begin{equation}
k_2 = \left[4 \mu_2^2 \over  \alpha_3 v_R^2\right] k_1 
\label{eq:k'}
\end{equation}

Ignoring the quantity in brackets of the first term of~(\ref{eq:higgskk'}) that has been fine tuned to be at weak scale,   note that $|\mu_2^2| <  \alpha_3 v_R^2 / 8$  must be satisfied  to get the right symmetry breaking pattern. Otherwise due to the $\mu_2^2$ term, there will be a saddle point either in the $k_2 = k_1$ or $k_2 = -k_1$ direction (depending on the sign of $\mu_2^2$), that can provide VEVs $\gtrsim v_R$ to $k_1$ and $k_2$, breaking $G_{LR}$ to $U(1)_{I_{3L+3R}} \times U(1)_{B-L}$ rather than to standard model.  Since we are not allowing any more fine tunings, the only way for $|\mu_2^2|$ to be smaller than $\alpha_3 v_R^2 / 8$ is to impose the $\mu-$symmetry, like we have done.

As long as the $\mu-$symmetry breaking is soft, $\mu_2^2$ does not receive quadratically divergent radiative correction from the Planck scale and can be naturally small.   However the $\mu-$symmetric Yukawa terms (with $\{Q_{iL}, \phi\} \rightarrow e^{i\beta} \{Q_{iL}, \phi\}$ and $Q_{iR}$ invariant under $\mu-$symmetry) which provide masses to quarks are of the form 
\begin{equation}
\label{eq:yuk}
\sum_{i,j = 1, 3}h_{ij}\bar{Q}_{iL} \phi Q_{jR} + h.c.
\end{equation}
where $Q_{iL} \equiv (u_{iL}, d_{iL})^T$ and $Q_{iR} \equiv (u_{iR}, d_{iR})^T$ are the left and right handed Quark doublets of the $i^{th}$ generation and are represented by $2 \times 1$ column vectors in iso-space.  Substituting for the VEVs of $\phi$ we can see that the up and down quark mass matrices are proportional to each other.  Therefore they can be simultaneously diagonalized and hereafter we work in the basis where $h_{ij}$ is diagonal. We can now obtain the quark masses in terms of the Yukawas.  Writing these explicitly we get,
\begin{equation}   
 {{h_{33}k_1} \over \sqrt{2}} = m_t, \ {{h_{33} k_2} \over {\sqrt{2}}} = m_b, \ {{h_{22} k_1}\over {\sqrt{2}}} = m_c 
\label{eq:masses}
\end{equation}
From the above we obtain $k_2/k_1 = m_b/m_t$.  It is useful to note here that the smallness of bottom to top quark mass ratio is natural in the standard model due to its approximate chiral symmetries that make small fermion Yukawa couplings like $h_b << h_t \sim 1$ of bottom quark natural. However in the LR model the same Yukawa coupling $h_{33}$ of equation~(\ref{eq:yuk}) provides both the top and bottom quarks their masses, and it is the chiral $\mu-$symmetry which keeps $\mu_2^2$ naturally small, that through equation~(\ref{eq:k'}) makes smallness of $k_2/k_1$ and therefore of $m_b/m_t$ natural. If we allow fine tuning of $\mu_2^2$   we will lose the naturalness of small fermion mass ratios like $m_b/m_t$ in the LR model.  %Therefore this symmetry is the generalization of the chiral symmetry of the standard model that makes the small fermion mass ratio $m_b/m_t$ natural. 

However the mass of the strange quark turns out to be too low, $m_s = h_{22} k_2 / \sqrt{2} = (m_b/m_t) m_c$.   Also the Cabibo-Kobayashi-Maskawa (CKM) mixing angles vanish. Therefore we must allow for approximate breaking of the $\mu-$symmetry by Yukawa terms of the kind
\begin{equation}
\tilde{h}_{22} \bar{Q}_{2L} \tilde{\phi}Q_{2R} +   \tilde{h}_{23} \bar{Q}_{2L} \tilde{\phi}Q_{3R}  + \\ \tilde{h}_{23}^{\star} \bar{Q}_{3L} \tilde{\phi}Q_{2R} + h.c. + ...
\label{eq:yukawabkg}
\end{equation}
so that we now have 
\begin{equation}
m_s \approx {{\tilde{h}_{22} k_1} \over {\sqrt {2}}},  \ V_{ts} \approx {{\tilde{h}_{23}k_1} \over {\sqrt{2}m_b}}
\label{eq:strangemass}
\end{equation}

Note that the occurrence of $\tilde{h}_{23}^\star (=\tilde{h}_{32})$ in equation~(\ref{eq:yukawabkg}) is because Yukawa matrices involving the bi-doublet are Hermitian due to parity, as is well known in left-right symmetric models~\cite{Duka:1999uc}.

\begin{figure}[ht]
\begin{center}
\includegraphics[width=4.5cm]{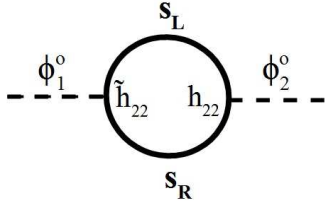}
\end{center}
\caption{Quadratically divergent radiative contribution to $\mu_2^2$ from approximate breaking of $\mu-$symmetry. $\phi^o_1$ and $\phi^o_2$ are neutral components of $\phi$ and are given in equation~(\ref{eq:fields}) }
\label{fig:oneloopmass}
\end{figure}

Since $\mu-$symmetry is approximately broken by Yukawa terms, $\mu_2^2$ receives a quadratically divergent radiative contribution at 1-loop level from diagrams involving the second generation such as the one  in Figure~\ref{fig:oneloopmass}.

%\end{center}
We evaluate such diagrams by providing a cut-off at the reduced Planck scale and find  the radiative correction at one-loop level,
\begin{equation}
\mu_2^{2(rad)} \sim \left({6\over 2}\right)\left[{{M_{Pl}^2}\over {16 \pi^2}}\right] h_{22} \tilde{h}_{22}
\label{eq:murad}
\end{equation}
The factor of 6 in the first numerator is to account for the 3 colors of the strange quark, and because there is also an equal contribution from a similar diagram with the charm quark in the loop (with $h_{22}$ and $\tilde{h}_{22}$ interchanged). The 2 in the denominator accounts for the trace in eq.~(\ref{eq:vbreak}).    

Since
\begin{equation}
\mu_2^2 = \mu_2^{2(bare)} + \mu_2^{2(rad)},
\label{eq:bare}
\end{equation}
if we do not allow fine tuning of equation~(\ref{eq:bare}), so that there is no precise cancellation between the bare and radiative terms, we obtain the bound
\begin{equation}
 \left| \mu_2^2 \right| \gtrsim \left| \mu_2^{2(rad)} \right| 
\label{eq:mu2bound}
\end{equation}
This bound on $\mu_2^2$ translates to a bound on $v_R^2$ through equation~(\ref{eq:k'}). Combining equations~(\ref{eq:mu2bound}),~(\ref{eq:murad}) and~(\ref{eq:k'}), and using~(\ref{eq:masses}) and~(\ref{eq:strangemass}) to express the Yukawas in terms of quark masses (with $k_1 \approx v_{wk}$), we get the following lower bound on $v_R$
\begin{equation}
v_R \sim v_{B-L} \gtrsim \left[{M_{Pl} \over {2\pi}}\right] \left[{\sqrt{6 m_s m_c \over \alpha_3 m_b m_t}}\right] \left[{m_t \over v_{wk}}\right]
\label{eq:bound}
\end{equation}

Several papers have results on the running Yukawa couplings in the standard model.  We use the results of Das and Parida~\cite{Das:2000uk} (updated recently in~\cite{Bora:2012tx}) for the quark masses in the standard model evaluated at a scale of $2 \times 10^{16} GeV$ to evaluate the above. From their work we have at the high scale, $m_s = 20.4 MeV, m_c = 0.22 GeV, m_b = 0.93 GeV, m_t = 70.5 GeV$ and $v_{wk} = 155 GeV \sqrt{2} \approx 219 GeV.$ 

Substituting in~(\ref{eq:bound}),  we get the bound 
\begin{equation}
\sqrt{\alpha_3}v_{R} \gtrsim 1.04 \times 10^{-3} M_{Pl} \sim 2.5 \times 10^{15}GeV.  
\label{eq:boundvalue}
\end{equation}
where we used $M_{Pl} = 2.4 \times 10^{18} GeV$. Note that the left hand side of~(\ref{eq:boundvalue}) is $\sqrt{2}$ times the mass of the second Higgs doublet.  Since we expect in the perturbative regime, $\alpha_3 \lesssim 1,$ equation~(\ref{eq:boundvalue}) evaluated with $\alpha_3 =1$ also provides a lower bound on $v_R$. That is
%\begin{equation}
$v_{R} \gtrsim  2.5 \times 10^{15}GeV.$  
%\label{eq:boundvaluevR}
%\end{equation}

Note that instead of the scale $2 \times 10^{16} GeV$, if we use the Yukawa couplings evaluated at the scale $10 TeV$ (from a recent paper by Antusch and Maurer~\cite{Antusch:2013jca}), we get $\sqrt{\alpha_3} v_R \geq 1.8 \times 10^{-3} M_{Pl}$.  This shows that the scale or the method used to evaluate the Yukawas does not make much of a difference to the bound.

The dots in equation~(\ref{eq:yukawabkg}) represent other terms that approximately violate $\mu-$symmetry, that can potentially be there.  However for the purposes of our lower bound calculation, the strength of their couplings can be neglected without resorting to fine-tuning.  For example  the term $\tilde{h}_{33} \bar{Q}_L \tilde{\phi} Q_R$, can be radiatively generated at one-loop with the strength (up to logarithmic factors) $\tilde{h}^{rad}_{33} \sim [1/(16\pi^2)](\tilde{h}_{32}\tilde{h}_{22} \tilde{h}_{23}) \sim 10^{-13}$. Since $h_{ij}$ is diagonal there is no contribution to $\tilde{h}^{rad}_{33}$ from the combination of Yukawas $(h_{33} \tilde{h}_{32} h_{23})$. %Thus $\tilde{h}_{33}$ while being larger than $10^{-13}$ to avoid fine tuning, can be chosen small enough so that it's contribution to equation~(\ref{eq:murad}) (with subscripts $22 \rightarrow 33$) can be ignored.  
Therefore we can choose $\tilde{h}_{33} \lesssim h_{22} \tilde{h}_{22}$ without fine tuning, so that its contribution to $\mu_2^{2(rad)}$ is at most of the same order of magnitude as already present in equation~(\ref{eq:murad}). 

\section{Seesaw scale}
%\emph{Seesaw scale --} 
The bound value $v_R \sim 2.5 \times 10^{15}GeV$ we obtained is close to the seesaw scale hinted by neutrino experiments namely $M_{ss} \sim v^2_{wk}/\sqrt{|\Delta m^2_{23}|} \sim 1.3 \times 10^{15} GeV$, which suggests that the seesaw scale is determined by the hierarchy problem and small quark mass ratios.  Since we have imposed an approximate $\mu-$symmetry, we will now verify that the terms necessary for seesaw mechanism are not suppressed.

As before under $\mu-$symmetry, $\{\phi, Q_{iL}, L_{iL}\} \rightarrow e^{i\beta} \{ \phi, Q_{iL},  L_{iL} \}$, and  $\Delta_L \rightarrow e^{-2i\beta} \Delta_L$, with $\Delta_R, Q_{iR}$ and $L_{iR}$ being invariant. There is only one  $P$ and $\mu-$symmetric quartic term in the Higgs potential that contains all three fields (as before, we use the standard notation in~\cite{Duka:1999uc,PhysRevD.65.095003}),
\begin{equation}
\beta_2 \left[Tr\left(\tilde{\phi} \Delta_R \phi^\dagger \Delta_L^\dagger\right) + Tr\left(\tilde{\phi}^\dagger \Delta_L \phi \Delta_R^\dagger\right) \right].
\end{equation}
The remaining $\mu-$symmetric terms containing $\Delta_L$ relevant for providing it a VEV are
\begin{equation}
-\mu_3^2 Tr (\Delta_L \Delta_L^\dagger) + \rho_3 Tr(\Delta_L \Delta_L^\dagger) Tr(\Delta_R \Delta_R^\dagger)
\end{equation}
%Rewriting in terms of the VEVs using equation~(\ref{eq:vev}) we get the Higgs potential terms 
%$(\beta_2 /2) v_R v_L k_1^2 + [(\rho_3 v_R^2/4) -\mu_3^2/2] v_L^2$.  Minimizing with respect to $v_L$, using  $v_R^2 = \mu^2_3 / \rho_1$, and noting that the term in the square brackets must be positive
Rewriting the Higgs potential in terms of the VEVs using equation~(\ref{eq:vev}), recalling that $v_R^2 = \mu^2_3 / \rho_1$,  and minimizing with respect to $v_L$
 we get, $v_L \sim (\beta_2 / [\rho_3 -2 \rho_1]) (v_{wk}^2/ v_R)$, with $\rho_3 > 2\rho_1 > 0$. 

 Noting that the usual Yukawa terms that give rise to Majorana masses $f_{ij} L^T_{iL}\tau_2 \Delta_L L_{jL} + h.c.$ (and terms with subscript $L \rightarrow R$), are permitted by $\mu-$symmetry we find that $v_L$ contributes to the largest light neutrino mass $m^\nu_{i}$
\begin{equation} 
 m^\nu_{i} \sim {{\sqrt{2}f_{ii} v_L}} \sim  \sqrt{2}\left[{{f_{ii}\beta_2} \over {\left(\rho_3-2\rho_1\right)}}\right] \left({v_{wk}^2 \over v_R}\right)
\end{equation}
with $i=3$ (or $i=2$) depending on whether it is normal (or inverted) hierarchy. Substituting our bound value, $v_R \sim 2.5 \times 10^{15} GeV$ and $v_{wk} \sim 246 GeV$, we see that if the quantity in the square brackets has a natural value around $1.4$ for $i=3$ (or $i=2$), we obtain the observed $|m^{\nu2}_{3} - m^{\nu2}_{2}|  \sim |\Delta{m^2_{32}}| \sim 0.0023 eV^2$.    With $v_R \sim 2.5 \times 10^{15}GeV$, the full seesaw mechanism can proceed either as Type II~\cite{Magg:1980ut,Lazarides:1980nt,PhysRevD.23.165} or Type I~\cite{Minkowski:1977sc,van1979supergravity,Yanagida01091980,Mohapatra:1979ia} seesaw, or a hybrid of the two.

\section{A few comments}
%\noindent \emph{A few comments --}

\begin{itemize}
\item If  the hierarchy problem is solved by supersymmetry or another mechanism at a scale $\Lambda$, then $M_{Pl}$ in equations such as~(\ref{eq:murad}) and~(\ref{eq:bound}) will be replaced by $\Lambda$. %Supersymmetry, if it's there, must be broken at $M_{Pl}$ or $\Lambda$.

\item The lower end of the bound in equation~(\ref{eq:bound}), that is $v_R \sim 2.5 \times 10^{15} GeV$, corresponds to the minimum violation of $\mu-$symmetry needed to obtain the correct quark mass spectrum. It also corresponds to not introducing an additional scale through the term $\mu_2^2 Tr (\tilde{\phi}^\dagger \phi)$ -- that is if $\mu^{2(bare)}_2 = 0$ in equation~(\ref{eq:bare}) then only the radiative correction $\mu^{2(rad)}_2$ determines $\sqrt{\alpha_3}v_R$ to be at the lower end of the bound.

\item If we add a second bi-doublet $\phi'$, without any additional fine tuning, its mass will naturally be of the order $M_{\phi'}\gtrsim M_{Pl}/(4 \pi)$. Though with two bi-doublets the $\mu-$symmetry can be broken only softly, the soft symmetry breaking mass term involving $\phi'$ such as $Tr \tilde{\phi}^\dagger \phi'$ must be at a large scale $\sim (m_s /m_t) M^2_{\phi'}$ in order to obtain the needed VEVs for $\phi'$.  Thus the $\mu-$symmetry is once again broken at a large scale, which in turn radiatively generates the $\mu_2^2$ term of first bi-doublet $\phi$ and results in a significant lower bound on $v_R$ as before.  Depending on choice of parameters,  the exact bound value will change.  The effective theory below $M_{\phi'}$ is the minimal model.

\item $\alpha_2$, the only complex parameter in the Higgs potential of the left-right symmetric model (see for example~\cite{Duka:1999uc})  is naturally small, since the term $\alpha_2 Tr(\tilde{\phi}^\dagger \phi) Tr(\Delta_R^\dagger \Delta_R)$ (with its parity counterpart) breaks $\mu-$ symmetry. Thus this term is under control for the purposes of our calculation.  Choosing a natural value for $\mu-$breaking parameter $|\alpha_2| \lesssim h_{22} \tilde{h}_{22}$, it can be consistently ignored in our calculation. However in LR models it is the source of the strong $CP$ problem, which can be solved without requiring an axion as shown in~\cite{Kuchimanchi:2010xs,Kuchimanchi:2012xb}. If a family of vectorlike quarks needed for the strong $CP$ solution are at the Planck scale, the theory below it is our minimal left-right model. The predictions of this ultra-violet completion are a measurable strong $CP$ phase  (neutron electric dipole moment (EDM)) that is radiatively generated in a large region of the parameter space, no electron EDM due to the high scale of new physics; and in the minimal version an absence of all leptonic $CP$ phases~\cite{Kuchimanchi:2012xb,Kuchimanchi:2012te}. If $v_R \sim 10^{15} GeV$ is the next scale of new physics then these predictions maybe the only window for more evidence on left-right symmetry.
       
\end{itemize}

\section{Conclusion} 
In non supersymmetric theories fine-tuning of gauge symmetry breaking scales is necessary.  By imposing a very reasonable condition that only VEVs that are actually needed to break gauge symmetries are fine tuned, while any additional parameters including those that break chiral or $\mu-$symmetries are not fine tuned, we have shown that we can obtain meaningful lower bounds on some gauge symmetry breaking scales. In the absence of unnecessary fine tuning, we may need chiral symmetries to ensure that higher gauge symmetries break to the standard model rather than to some other group.  These chiral symmetries are however broken by fermion (quark) mass ratios.   The hierarchy problem then reappears to destabilize the symmetry breaking pattern, but is suppressed from the reduced Planck scale by small quark mass ratios that break certain chiral or $\mu-$symmetries. Since we do not allow an additional fine tuning relation, we can obtain a lower bound on where some gauge symmetries are broken.  

 In particular  we find that the $SU(2)_R \times U(1)_{B-L}$ breaking scale in minimal left right symmetric model, $v_{B-L} \sim v_R \gtrsim 10^{15}GeV$ if there is no fine tuning of parameters that break chiral or $\mu-$symmetry.  Since $B-L$ breaking triggers the seesaw mechanism, it leads to the understanding of seesaw scale $\sim 10^{15}GeV$ indicated by neutrino experiments, without need for grand unification constraints. 

Our bound can be violated in the minimal left-right symmetric model if there is an additional fine tuning relation involving parameters that break chiral symmetries -- apart from the fact that this is theoretically unattractive, it should be noted that in this case, the smallness of some fermion  mass ratios such as $m_b/m_t$ that is natural in the standard model,  would no longer be natural in the left-right symmetric model.

%\section*{References}
\bibliography{LRsymmetrythi}

\begin{thebibliography}{10}
\expandafter\ifx\csname url\endcsname\relax
  \def\url#1{\texttt{#1}}\fi
\expandafter\ifx\csname urlprefix\endcsname\relax\def\urlprefix{URL }\fi
\expandafter\ifx\csname href\endcsname\relax
  \def\href#1#2{#2} \def\path#1{#1}\fi

\bibitem{PhysRevD.14.1667}
E.~Gildener,
  \href{http://link.aps.org/doi/10.1103/PhysRevD.14.1667}{Gauge-symmetry
  hierarchies}, Phys. Rev. D 14 (1976) 1667--1672.
\newblock \href {http://dx.doi.org/10.1103/PhysRevD.14.1667}
  {\path{doi:10.1103/PhysRevD.14.1667}}.
\newline\urlprefix\url{http://link.aps.org/doi/10.1103/PhysRevD.14.1667}

\bibitem{PhysRevD.19.1277}
S.~Weinberg,
  \href{http://link.aps.org/doi/10.1103/PhysRevD.19.1277}{Implications of
  dynamical symmetry breaking: An addendum}, Phys. Rev. D 19 (1979) 1277--1280.
\newblock \href {http://dx.doi.org/10.1103/PhysRevD.19.1277}
  {\path{doi:10.1103/PhysRevD.19.1277}}.
\newline\urlprefix\url{http://link.aps.org/doi/10.1103/PhysRevD.19.1277}

\bibitem{PhysRevD.20.2619}
L.~Susskind, \href{http://link.aps.org/doi/10.1103/PhysRevD.20.2619}{Dynamics
  of spontaneous symmetry breaking in the weinberg-salam theory}, Phys. Rev. D
  20 (1979) 2619--2625.
\newblock \href {http://dx.doi.org/10.1103/PhysRevD.20.2619}
  {\path{doi:10.1103/PhysRevD.20.2619}}.
\newline\urlprefix\url{http://link.aps.org/doi/10.1103/PhysRevD.20.2619}

\bibitem{Arvanitaki:2013yja}
A.~Arvanitaki, M.~Baryakhtar, X.~Huang, K.~Van~Tilburg, G.~Villadoro, {The Last
  Vestiges of Naturalness}\href {http://arxiv.org/abs/1309.3568}
  {\path{arXiv:1309.3568}}.

\bibitem{Bertuzzo:2013dja}
E.~Bertuzzo, {SUSY after LHC8: a brief overview}, EPJ Web Conf. 60 (2013)
  18001.
\newblock \href {http://arxiv.org/abs/1307.0318} {\path{arXiv:1307.0318}},
  \href {http://dx.doi.org/10.1051/epjconf/20136018001}
  {\path{doi:10.1051/epjconf/20136018001}}.

\bibitem{Anton}
I.~Antoniadis,
  \href{http://vietnam.in2p3.fr/2013/Inauguration/transparencies/Antoniadis.pdf}{{Theory
  and phenomenology of physics with extra dimensions}}, IXth Rencontres du
  Vietnam: Windows on the Universe, Quy Nhon Vietnam.
\newline\urlprefix\url{http://vietnam.in2p3.fr/2013/Inauguration/transparencies/Antoniadis.pdf}

\bibitem{Beringer:1900zz}
J.~Beringer, et~al., {Review of Particle Physics (RPP)}, Phys.Rev. D86 (2012)
  010001.
\newblock \href {http://dx.doi.org/10.1103/PhysRevD.86.010001}
  {\path{doi:10.1103/PhysRevD.86.010001}}.

\bibitem{Minkowski:1977sc}
P.~Minkowski, {mu e gamma at a Rate of One Out of 1-Billion Muon Decays?},
  Phys.Lett. B67 (1977) 421.
\newblock \href {http://dx.doi.org/10.1016/0370-2693(77)90435-X}
  {\path{doi:10.1016/0370-2693(77)90435-X}}.

\bibitem{van1979supergravity}
M.~Gell-Mann, P.~Ramond, R.~Slansky,
  \href{http://books.google.co.in/books?id=Dg-CAAAAIAAJ}{Supergravity}, ed. by
  D. Freedman and P. Van Nieuwenhuizen, North Holland, Amsterdam, 315-321,
  1979.
\newline\urlprefix\url{http://books.google.co.in/books?id=Dg-CAAAAIAAJ}

\bibitem{Yanagida01091980}
T.~Yanagida, Horizontal symmetry and masses of neutrinos, Progress of
  Theoretical Physics 64~(3) (1980) 1103--1105.
\newblock \href {http://dx.doi.org/10.1143/PTP.64.1103}
  {\path{doi:10.1143/PTP.64.1103}}.

\bibitem{Mohapatra:1979ia}
R.~N. Mohapatra, G.~Senjanovic, {Neutrino Mass and Spontaneous Parity
  Violation}, Phys.Rev.Lett. 44 (1980) 912.
\newblock \href {http://dx.doi.org/10.1103/PhysRevLett.44.912}
  {\path{doi:10.1103/PhysRevLett.44.912}}.

\bibitem{Altarelli:2013aqa}
G.~Altarelli, D.~Meloni, {A non supersymmetric SO(10) grand unified model for
  all the physics below $M_(GUT)$}, JHEP 1308 (2013) 021.
\newblock \href {http://arxiv.org/abs/1305.1001} {\path{arXiv:1305.1001}},
  \href {http://dx.doi.org/10.1007/JHEP08(2013)021}
  {\path{doi:10.1007/JHEP08(2013)021}}.

\bibitem{delAguila:1980at}
F.~del Aguila, L.~E. Ibanez, {Higgs Bosons in SO(10) and Partial Unification},
  Nucl.Phys. B177 (1981) 60.
\newblock \href {http://dx.doi.org/10.1016/0550-3213(81)90266-2}
  {\path{doi:10.1016/0550-3213(81)90266-2}}.

\bibitem{Dimopoulos:1984ha}
S.~Dimopoulos, H.~Georgi, {Extended Survival Hypothesis and Fermion Masses},
  Phys.Lett. B140 (1984) 67.
\newblock \href {http://dx.doi.org/10.1016/0370-2693(84)91049-9}
  {\path{doi:10.1016/0370-2693(84)91049-9}}.

\bibitem{PhysRevD.27.1601}
R.~N. Mohapatra, G.~Senjanovi\ifmmode~\acute{c}\else \'{c}\fi{},
  \href{http://link.aps.org/doi/10.1103/PhysRevD.27.1601}{Higgs-boson effects
  in grand unified theories}, Phys. Rev. D 27 (1983) 1601--1612.
\newblock \href {http://dx.doi.org/10.1103/PhysRevD.27.1601}
  {\path{doi:10.1103/PhysRevD.27.1601}}.
\newline\urlprefix\url{http://link.aps.org/doi/10.1103/PhysRevD.27.1601}

\bibitem{PhysRevD.10.275}
J.~C. Pati, A.~Salam, Lepton number as the fourth "color", Phys. Rev. D 10~(1)
  (1974) 275--289.
\newblock \href {http://dx.doi.org/10.1103/PhysRevD.10.275}
  {\path{doi:10.1103/PhysRevD.10.275}}.

\bibitem{PhysRevD.11.566}
R.~N. Mohapatra, J.~C. Pati, Left-right gauge symmetry and an "isoconjugate"
  model of $cp$ violation, Phys. Rev. D 11~(3) (1975) 566--571.
\newblock \href {http://dx.doi.org/10.1103/PhysRevD.11.566}
  {\path{doi:10.1103/PhysRevD.11.566}}.

\bibitem{Senjanovic:1975rk}
G.~Senjanovic, R.~N. Mohapatra, {Exact Left-Right Symmetry and Spontaneous
  Violation of Parity}, Phys. Rev. D12 (1975) 1502.
\newblock \href {http://dx.doi.org/10.1103/PhysRevD.12.1502}
  {\path{doi:10.1103/PhysRevD.12.1502}}.

\bibitem{Duka:1999uc}
P.~Duka, J.~Gluza, M.~Zralek, {Quantization and renormalization of the manifest
  left- right symmetric model of electroweak interactions}, Annals Phys. 280
  (2000) 336--408.
\newblock \href {http://arxiv.org/abs/hep-ph/9910279}
  {\path{arXiv:hep-ph/9910279}}, \href
  {http://dx.doi.org/10.1006/aphy.1999.5988}
  {\path{doi:10.1006/aphy.1999.5988}}.

\bibitem{PhysRevD.65.095003}
G.~Barenboim, M.~Gorbahn, U.~Nierste, M.~Raidal,
  \href{http://link.aps.org/doi/10.1103/PhysRevD.65.095003}{Higgs sector of the
  minimal left-right symmetric model}, Phys. Rev. D 65 (2002) 095003.
\newblock \href {http://dx.doi.org/10.1103/PhysRevD.65.095003}
  {\path{doi:10.1103/PhysRevD.65.095003}}.
\newline\urlprefix\url{http://link.aps.org/doi/10.1103/PhysRevD.65.095003}

\bibitem{Vissani:1997ys}
F.~Vissani, {Do experiments suggest a hierarchy problem?}, Phys.Rev. D57 (1998)
  7027--7030.
\newblock \href {http://arxiv.org/abs/hep-ph/9709409}
  {\path{arXiv:hep-ph/9709409}}, \href
  {http://dx.doi.org/10.1103/PhysRevD.57.7027}
  {\path{doi:10.1103/PhysRevD.57.7027}}.

\bibitem{Casas:2004gh}
J.~Casas, J.~Espinosa, I.~Hidalgo, {Implications for new physics from
  fine-tuning arguments. 1. Application to SUSY and seesaw cases}, JHEP 0411
  (2004) 057.
\newblock \href {http://arxiv.org/abs/hep-ph/0410298}
  {\path{arXiv:hep-ph/0410298}}, \href
  {http://dx.doi.org/10.1088/1126-6708/2004/11/057}
  {\path{doi:10.1088/1126-6708/2004/11/057}}.

\bibitem{Das:2000uk}
C.~Das, M.~Parida, {New formulas and predictions for running fermion masses at
  higher scales in SM, 2 HDM, and MSSM}, Eur.Phys.J. C20 (2001) 121--137.
\newblock \href {http://arxiv.org/abs/hep-ph/0010004}
  {\path{arXiv:hep-ph/0010004}}, \href
  {http://dx.doi.org/10.1007/s100520100628} {\path{doi:10.1007/s100520100628}}.

\bibitem{Bora:2012tx}
K.~Bora, {Updated values of running quark and lepton masses at GUT scale in SM,
  2HDM and MSSM}\href {http://arxiv.org/abs/1206.5909}
  {\path{arXiv:1206.5909}}.

\bibitem{Antusch:2013jca}
S.~Antusch, V.~Maurer, {Running quark and lepton parameters at various scales},
  JHEP 1311 (2013) 115.
\newblock \href {http://arxiv.org/abs/1306.6879} {\path{arXiv:1306.6879}},
  \href {http://dx.doi.org/10.1007/JHEP11(2013)115}
  {\path{doi:10.1007/JHEP11(2013)115}}.

\bibitem{Magg:1980ut}
M.~Magg, C.~Wetterich, Phys.Lett. B94 (1980) 61.

\bibitem{Lazarides:1980nt}
G.~Lazarides, Q.~Shafi, C.~Wetterich, {Proton Lifetime and Fermion Masses in an
  SO(10) Model}, Nucl.Phys. B181 (1981) 287--300.
\newblock \href {http://dx.doi.org/10.1016/0550-3213(81)90354-0}
  {\path{doi:10.1016/0550-3213(81)90354-0}}.

\bibitem{PhysRevD.23.165}
R.~N. Mohapatra, G.~Senjanovi\ifmmode~\acute{c}\else \'{c}\fi{},
  \href{http://link.aps.org/doi/10.1103/PhysRevD.23.165}{Neutrino masses and
  mixings in gauge models with spontaneous parity violation}, Phys. Rev. D 23
  (1981) 165--180.
\newblock \href {http://dx.doi.org/10.1103/PhysRevD.23.165}
  {\path{doi:10.1103/PhysRevD.23.165}}.
\newline\urlprefix\url{http://link.aps.org/doi/10.1103/PhysRevD.23.165}

\bibitem{Kuchimanchi:2010xs}
R.~Kuchimanchi, {$P/CP$ Conserving $CP/P$ Violation Solves Strong $CP$
  Problem}, Phys. Rev. D82 (2010) 116008.
\newblock \href {http://arxiv.org/abs/1009.5961} {\path{arXiv:1009.5961}},
  \href {http://dx.doi.org/10.1103/PhysRevD.82.116008}
  {\path{doi:10.1103/PhysRevD.82.116008}}.

\bibitem{Kuchimanchi:2012xb}
R.~Kuchimanchi, {Maximal $CP$ and Bounds on the Neutron Electric Dipole Moment
  from $P$ and $CP$ Breaking}, Phys.Rev. D86 (2012) 036002.
\newblock \href {http://arxiv.org/abs/1203.2772} {\path{arXiv:1203.2772}},
  \href {http://dx.doi.org/10.1103/PhysRevD.86.036002}
  {\path{doi:10.1103/PhysRevD.86.036002}}.

\bibitem{Kuchimanchi:2012te}
R.~Kuchimanchi, \href{http://dx.doi.org/10.1140/epjc/s10052-014-2726-5}{P
  stabilizes dark matter and with cp can predict leptonic phases}, Eur. Phys.
  J. C 74~(2) (2014) 1--10.
\newblock \href {http://arxiv.org/abs/1209.3031} {\path{arXiv:1209.3031}},
  \href {http://dx.doi.org/10.1140/epjc/s10052-014-2726-5}
  {\path{doi:10.1140/epjc/s10052-014-2726-5}}.
\newline\urlprefix\url{http://dx.doi.org/10.1140/epjc/s10052-014-2726-5}

\end{thebibliography}

\end{document}